\documentclass{osa-article}

\journal{oe}

\usepackage[normalem]{ulem}

\definecolor{vert}{rgb}{0.01, 0.75, 0.24}
\definecolor{red}{rgb}{1, 0, 0}

\begin{document}

\title{Single-shot hybrid photoacoustic-fluorescent microendoscopy through a multi-mode fiber with wavefront shaping}

\author{Sylvain Mezil\authormark{1}, Antonio M. Caravaca-Aguirre\authormark{1}, Edward Z. Zhang\authormark{2}, Philippe Moreau\authormark{1}, Ir\`ene Wang\authormark{1}, Paul C. Beard\authormark{2} and Emmanuel Bossy\authormark{1,*}}

\address{\authormark{1}Univ. Grenoble Alpes, CNRS, LIPhy, 38000 Grenoble, France\\
\authormark{2}Department of Medical Physics and Biomedical Engineering, University College London, Gower Street, London WC1E 6BT, UK}

\email{\authormark{*}emmanuel.bossy@univ-grenoble-alpes.fr} 



\begin{abstract}
We present a minimally-invasive endoscope based on a multimode fiber that combines photoacoustic and fluorescence sensing. From the measurement of a transmission matrix during a prior calibration step, a focused spot is produced and raster-scanned over a sample at the distal tip of the fiber by use of a fast spatial light modulator. 
An ultra-sensitive fiber-optic ultrasound sensor for photoacoustic detection placed next to the fiber is combined with  a photodetector to obtain 
both fluorescence and photoacoustic images with a distal imaging tip no larger than 250~\textmu m. 
The high signal-to-noise ratio provided by wavefront shaping based focusing and the ultra-sensitive ultrasound sensor enables imaging with a single laser shot per pixel, demonstrating fast   two-dimensional hybrid \textit{in vitro} imaging of  red blood cells and fluorescent beads. 
\end{abstract}

\section{Introduction}
Photoacoustic imaging is a promising modality for numerous biomedical applications\cite{Wang2016}. It provides a contrast based on optical absorption and is either used as a label-free method to image endogenous light absorbing molecules in tissues, such as hemoglobin, melanin and lipids, or combined with exogenous contrast agents to target specific structures. When absorbers are illuminated by light of time-varying intensity (such as provided by pulsed or modulated laser sources), photoacoustic waves are generated through thermoelasticity. The photoacoustic waves can then be detected by ultrasound detectors outside the sample. This modality may also provides depth information for 3D imaging. Two regimes can be distinguished for photoacoustic imaging: in the first, limited to superficial depth because of optical scattering, a focused laser beam is scanned within the sample and an image is formed point-by-point with optical resolution; in the second, the sample is excited with wide-field illumination and an image is reconstructed deep in tissue with acoustic resolution from the detection of unscattered photoacoustic waves at multiple locations. For both regimes, the spatial resolution of photoacoustic images decreases with increasing depth, due to light scattering for the optical-resolution regime, or sound attenuation for the acoustic resolution regime. In photoacoustic imaging, the depth-to-resolution ratio is limited to around 200, e.g., a 1~\textmu m resolution image can only be obtained at depth as shallow as 200~\textmu m. 

In this context, photoacoustic endoscopy appears as an alternative approach to retain high imaging resolution at depth, at the cost of invasiveness which should thus be minimized.
A number of photoacoustic endoscopic probes have been reported providing either acoustic-resolution\cite{Yang2012,yang14,Yang2011,Ji2015,Hui2017} or optical-resolution \cite{hajireza11b,hajireza13,Bai2014,Dong2014,Guo2017,Liu2018,xi13,yang15} imaging. They generally include a fiber to bring the laser light to the imaging tip, a scanning mirror, an ultrasonic transducer, and, in the case of optical-resolution systems, a graded-index (GRIN) lens and/or fiber bundle to focus the excitation light. The footprint of such probes is usually in the millimeter range. Although this is adequate for a number of biomedical applications (such as intravascular\cite{Bai2014,Ji2015,Hui2017}, gastrointestinal tract\cite{Yang2012,yang14} or ovarian\cite{Yang2011} imaging), the insertion of a mm-sized probe would induce detrimental tissue damage in other cases, e.g., when studying neuro-vascular activity in the brain of small animals. In such cases, reducing the footprint of photoacoustic endoscope probes is an important requirement.

Fiber bundles are most commonly used to performed endoscopy with optical resolution, but their footprint remain on the order of a millimeter. For an equivalent information content, multimode fibers (MMF) have a much smaller diameter (typically 20 to 30 times smaller in area) than fiber bundles and  have been implemented as ultrathin endoscope probes in various all-optical modalities, such as reflection\cite{choi12,mahalati13} and confocal\cite{Loterie2015} microscopy, optical trapping\cite{Leite2018}, fluorescence\cite{cizmar12,Papadopoulos2013,Caravaca-Aguirre2013,caravaca17,ohayon18,turtaev18,vasquez18}, Raman\cite{Gusachenko2017}, two-photon\cite{Morales-Delgado2015} and light-sheet\cite{ploshner15} microscopy. MMF have also been used in photoacoustic imaging to deliver excitation light, combined with an external ultrasound transducer\cite{papadopoulos13, simandoux15}. The guided modes in MMF have different phase velocities, and consequently coherent light injected into a MMF most generally emerges as speckle patterns. Although speckle illumination may be exploited directly in some cases~\cite{mahalati13,caravaca19}, most applications require to image the sample by scanning a focused excitation spot at the MMF output, which can be obtained by wavefront shaping. Initially developed to focus light through scattering media\cite{Vellekoop07}, wavefront shaping consists in manipulating the complex wavefront injected into a MMF in order to obtain a given output field pattern. It is based on the existence of a linear and deterministic relation between incident and transmitted fields, which needs to be evaluated. Several techniques has been employed to implement wavefront shaping, such as iterative optimization\cite{dileonardo11,Cizmar2011,Gu2014}, digital phase conjugation\cite{papadopoulos12,Morales-Delgado2015}, or transmission matrix measurement\cite{choi12,Ploeschner2014,Loterie2015}. The input wavefront phase modulation may be performed using either liquid-crystal spatial light modulators (LC-SLM) or, more recently, digital micro-mirror devices (DMD)\cite{Caravaca-Aguirre2013}, which have the advantage of speed but at the cost of light efficiency and are binary amplitude modulation devices\cite{Turtaev2017}.  

Recently, ultrathin optical-resolution photoacoustic endoscopy has been demonstrated with
both capillary waveguide~\cite{stasio15} and a MMF~\cite{caravaca19}. In Ref.~\cite{stasio15},  the excitation light guided through a 350~\textmu m-wide 
silica capillary tube is focused by digital phase conjugation at the endoscope distal tip, while the capillary core, filled with water, carries back the acoustic waves towards the endoscope input where a transducer is placed. However, only relatively large objects (such as nylon threads) have been imaged, and the length of the device is limited to a few cm by sound attenuation in the water core. In Ref.~\cite{caravaca19}, a hybrid photoacoustic-fluorescence based on a MMF for light excitation and a fiber ultrasound sensor for acoustic detection was presented. The proposed approach involved illuminating the sample with a set of `natural' speckles obtained at the MMF output by projecting random intensity patterns at its input, using a digital mirror device (DMD). Photoacoustic and fluorescence signals are detected for each of these known speckles and an image is reconstructed by solving an inverse problem using a sparsity constraint. This strategy is simple to implement and can be combined with compressive sensing to reduce the number of measurements relative to the number of pixels in the final image. However, since the excitation power is spread over the whole field of view, the detected signals have a low signal-to-noise ratio (SNR) and a large number of laser pulses need to be averaged, which limits the imaging speed. In addition, the possibility to use compressive sensing methods is limited to sparse samples.

Previous ultrathin photoacoustic endoscopes presented relatively low SNR, requiring the acoustic signals to be averaged over many laser pulses, at the cost of long acquisition times and consequently a very low temporal resolution. Here we present an alternative MMF-based photoacoustic endoscopy approach, to improve measurement SNR and imaging speed. Using wavefront shaping, the excitation laser is focused at the distal tip of the MMF, which enhances the power density by several orders of magnitude, compared to speckle illumination~\cite{caravaca19}. Such focus can then be raster-scanned to image the sample.
The transmission matrix (TM) method is used to determine the MMF light-transport properties (as opposed to digital phase conjugation in Ref.~\cite{stasio15}): this technique is both easy to implement and versatile since, once measured, the TM enables one to create any pattern at the fiber output. 
Another crucial aspect is the acoustic detection: we use a fiber-optic ultrasound sensor (FOS) that consists of a single-mode optical fiber containing a polymer microcavity, acting as a Fabry-P\'erot interferometer, at its tip. Such a FOS is particularly attractive for our approach as its footprint is the one of a single-mode fiber. As the FOS and the MMF are set next to each other with the sample close to the MMF tip, the acoustic waves propagated from the sample to the FOS with large angles. 
Compared to the commercial FOS used in Ref.~\cite{caravaca19}, the novel FOS used here relies  on a plano-concave polymer microresonator and offers a much higher sensitivity and much larger angular acceptance over its whole bandwidth.

In this work, we demonstrate that, by combining wavefront-shaping to focus through a MMF and an ultra-sensitive FOS, photoacoustic imaging can be performed with a single laser pulse per pixel, through a 250$\times$125~\textmu m$^2$ footprint endoscope. This is an important step towards fast photoacoustic endoscopy. In addition, we show that fluorescence imaging can be performed with the exact same setup enhanced with a photomultiplier for fluorescence detection. 
Combining photoacoustic and fluorescence imaging into one single device is particularly attractive to offer a flexible imaging tool sensitive to different optical properties. For instance, vascular dynamics can be measured using photoacoustic imaging while neuronal activity can be monitored using the fluorescence signal emitted by calcium indicators.
As a proof-of-concept, this bimodal endoscope is used here to image red blood cells and fluorescent beads.

\section{Experimental methods}

\subsection{Hybrid photoacoustic-fluorescence probe}
The endoscope probe consists of a multimode fiber (MMF), used to both guide the excitation light and collect fluorescence signal, and a fiber-optic ultrasound sensor (FOS),  placed side-by-side, used to measure the photoacoustic signals from the sample. The MMF and the FOS are held together inside a cannula with about $\sim$15~mm of both fibers emerging out of the cannula . Its associated 
total footprint is 250$\times$125~\textmu m$^2$, as shown in Fig.~\ref{fig:setup}(b-c). The MMF is a $\sim$8~cm-long segment of  62.5~\textmu m-core and 125~\textmu m-cladding graded-index (GRIN) fiber.  GRIN fibers, due to weaker mode coupling compared to  step-index fibers, exhibit light transport properties that are more robust to deformation/bending\cite{caravaca17,BoonzajerFlaes2018}. The focus created through such a fiber is thus expected to be more stable.

The FOS is based on a single mode fiber (10~\textmu m core diameter and 125~\textmu m cladding) with a plano-concave polymer microresonator on its tip~\cite{guggenheim17}. 
Acoustic waves incident on the cavity modulate its optical thickness and thus its reflectivity which is detected using an interrogation laser beam coupled into the fibre. The sensor provides a wide acoustic bandwidth of 1-40MHz.
This new design offers a strong optical confinement which leads, compared to a planar cavity sensor, to a significantly larger sensitivity as well as wider directivity~\cite{morris09, guggenheim17}.
For a 65~\textmu m-wide sample located 100~\textmu m from the endoscope tip and aligned with the MMF axis (as in the following experiments), acoustic waves are incident on the microcavity with angles up to $\sim$60$^{\circ}$ from normal incidence. This novel plano-concave cavity exhibits effective omnidirectionnality with high sensitivity up to $\pm$90$^{\circ}$  over the whole 1-40~MHz bandwidth, while it is non-uniform for frequencies above 10~MHz for a planar cavity sensor (with drops up to 25~dB)~\cite{morris09}. 
The combination of high sensitivity and omnidirectionality provided by the plano-concave cavity design results in high acoustic SNR thus enabling single-shot photoacoustic detection at each illumination point without signal averaging.

\subsection{\label{sec:setup}Experimental setup}

\begin{figure}[t]
\centering
\includegraphics[width=\linewidth]{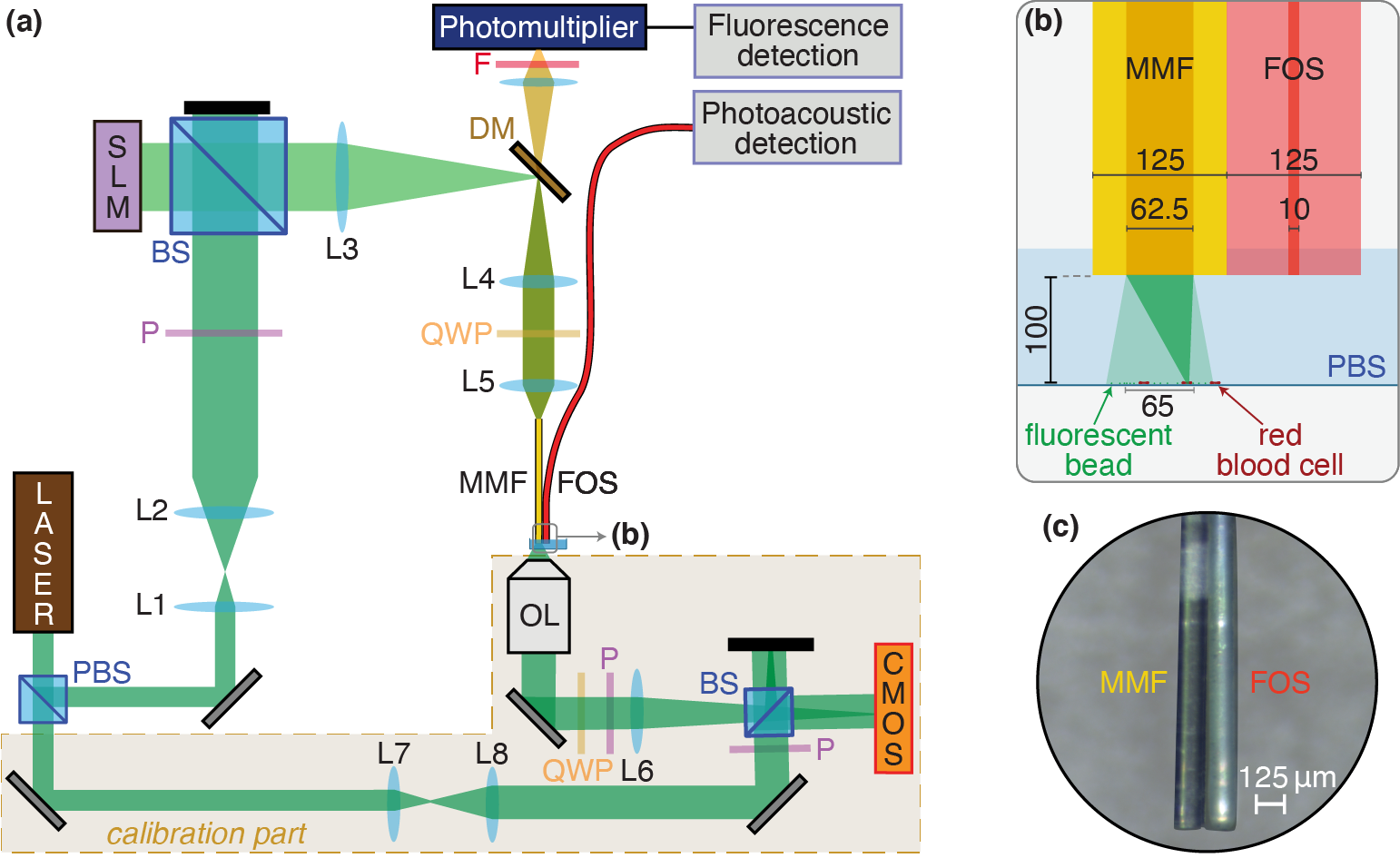}
\caption{(a)~Experimental setup. SLM: Spatial Light Modulator, (P)BS: (Polarizing) Beam Splitter. L: Lens, P: Polarizer, QWP: Quarter Wave Plate. F: Filter. MMF: Multimode Fiber. FOS: Fiber-Optic ultrasound sensor. DM: Dichroic Mirror. (b)~Zoom of the setup at the endoscope tip in presence of the sample. PBS: Phosphate-buffered saline. Dimensions in micrometer. (c)~Photography of the endoscope tip.} 
\label{fig:setup}
\end{figure}

The experimental setup is presented in Fig.~\ref{fig:setup}(a). Light pulses delivered by a $\lambda=532$~nm  Q-switched laser (Cobolt Tor 400) with a repetition frequency of $f_\text{rep}=7$~kHz and a pulse duration of $\sim$5~ns, is divided into two beams. One beam is used as the reference beam for calibration (see `calibration part' in Fig.~\ref{fig:setup}(a)). The other beam provides the excitation light for photoacoustic and fluorescence imaging. The excitation beam is first expanded (through $L1$ and $L2$ lenses, Fig.~\ref{fig:setup}) and sent to a fast LC-SLM for phase modulation (Meadowlark HSP512L-532, 512$\times$512 pixels, 380 Hz refreshment rate) 
at normal incidence by a beamsplitter cube (50:50). The beam transmitted by the cube
is reflected by a dichroic mirror. Then, after going through a telescope ($f_{L3}$=100~mm and $f_{L4}$=50~mm-focal length achromatic doublets), the beam is circularly polarized by a quarter-wave plate, as circular polarization is well-preserved when propagating through a straight fiber. It is then focused by an aspheric lens (Thorlabs AL12106-A, $f_{L5}$=10 mm) into a $\sim$8~cm-long section of multimode fiber (Thorlabs GIF625-10). The output facet of the MMF and the FOS tip, held next to each other, are immersed in water or phosphate-buffered saline (PBS), and the sample surface is placed at a distance of $\sim$100~\textmu m below the probe. This distance is arbitrarily chosen as a good compromise between reducing the angle between the acoustic waves generated from the RBCs and the axis of the FOS, while being close enough to maintain a high numerical aperture (NA) to achieve good focusing (see Sec.~\ref{sec:focus}). 

The sample consists of red blood cells (RBCs) and fluorescent beads (1~\textmu m diameter, Invitrogen F8851) in a glass-bottom well and is prepared in the following way. First, an aqueous solution of fluorescent beads is deposited on the glass slide at the bottom of a clean well. After a few minutes, the solution is air-dried leaving only the beads adsorbed on the glass surface. Then the solution of RBCs (blood diluted $\sim$6000 times) in PBS is deposited in the well and the RBCs are left to sediment before acquisition. Hence, both fluorescent beads and RBCs are observed on the bottom surface of the well (see Fig.~\ref{fig:setup}(b)).

Fluorescence light emitted by the beads is collected by the MMF and travels back along the same path until it is transmitted by the dichroic mirror and detected by a photomultiplier tube (PMT, Hamamatsu R647) through a long-pass filter (Chroma ET575lp). The signal from the PMT is then sent to a current preamplifier (Stanford Research Systems SR570) before acquisition. The photoacoustic signal detection is performed as detailed in Ref.~\cite{guggenheim17}. The signals from the PMT and the FOS  are then acquired by a USB oscilloscope (TiePie Handyscope HS6) on a computer. 

In order to obtain the transmission matrix to further perform wavefront shaping, one needs to measure the complex optical field at the MMF output. The amplitude and phase of the speckle in the output plane can be extracted through off-axis holography~\cite{cuche00}, i.e., by measuring the interference between the fiber output field and a reference beam. To that end, the output field of the MMF is imaged by an objective lens (20x Mitutoyo Plan Apochromat, NA=0.42) and an achromatic doublet ($f_{L6}$=200~mm). Its circular polarization is converted back into linear polarization.
Then it is merged with the reference beam by a non-polarizing beam-splitter. The interference intensity pattern is recorded by a CMOS camera (Basler acA1300-200um, see `calibration part' in Fig.~\ref{fig:setup}(a)). The output field is retrieved, in amplitude and phase, from the interferogram in the Fourier domain by extracting the spatial frequencies around the first diffraction order. During the calibration procedure, the sample is translated so that the output light from the MMF illuminates an area without beads nor RBCs (right side of the sample in Fig.~\ref{fig:setup}(b)).

\subsection{\label{sec:calibration}Calibration procedure}
The MMF is calibrated using the transmission matrix (TM) approach\cite{Popoff10a}. It consists in measuring the deterministic relation that connects the optical field in some imaging plane at the distal tip of the MMF (expressed in a set of output modes) to the input field at the proximal tip, decomposed on a set of input modes. In our case, the input modes are defined as plane waves in the SLM plane with various $\vec{k}$ directions. 
The set of input modes in the SLM plane corresponds to focused spots at the input facet of the MMF, defined across a 60$\times$60 orthogonal grid (with 1.17~\textmu m spacing). 
Retaining only the input spots that fall onto the MMF core reduces the number of effective input modes to approximately $N_{in}=2500$. 
This roughly corresponds to the number of guided modes in the GRIN MMF if we consider in first approximation this number to be $N_\text{MMF}=(1/4)\cdot(\pi  D \text{NA} / \lambda)^2=2600$, with $D$ the MMF core diameter (62.5~\textmu m). The output modes are defined over a grid, with $\sim$1.3~\textmu m spacing in the imaging plane, which is set in a plane 100~\textmu m far from the fiber distal end (named `output plane' in the following). 
The image field of view is limited to a circular region of $\sim$65~\textmu m diameter, leading to a number of output modes of about $N_{out}=1800$. The size of this region was chosen to be small enough to limit the TM acquisition time and calculation duration and ensure a good focusing quality (see Sec.~\ref{sec:focus}). Prior to the acquisition of the transmission matrix, the MMF input facet is centered by an automatic procedure, consisting of iteratively scanning a focused spot in two orthogonal directions of the MMF entrance plane, using the SLM, while monitoring the integrated intensity at the MMF output. The input modes are then defined around this center position. Note that the MMF is positioned so that the specular reflection from the SLM, corresponding to unmodulated light, is focused outside the fiber core and is not guided to the sample, i.e., the MMF axis is slightly shifted from the main optical axis.

Each column of the transmission matrix $K$ is acquired by applying on the SLM a linear phase ramp to produce a plane wave with a given $\vec{k}$, corresponding to an input mode, and by extracting the associated amplitude and phase in the output plane through off-axis holography (see Sec.~\ref{sec:setup})~\cite{cuche00}.
The complex optical field values at each of the pixels defined as output modes then form one column of the TM.
Once  this is repeated for all input modes, the TM acquisition is completed, and the relation between the input field $E_{in}$ and the output one $E_{out}$ can be expressed as $E_{out}=K\cdot E_{in}$.
Assuming
\begin{equation}
    E_{in}=K^\dagger\cdot E_{out},
    \label{eq:TM}
\end{equation}
where $^\dagger$ represents the transpose conjugate~\cite{Popoff10a}, one can design input fields to obtain any desired output pattern. To raster scan the sample, one needs to generate individual output modes (focused spots). In this case, $E_{out}$ is a basis vector with only one non-zero element. The computed input field, which is combination of the input modes (in the tilted plane waves basis), are defined in amplitude and phase, but only the phase patterns are stored to be displayed on the SLM, since the latter is a phase-only modulator. 
This entire calibration procedure is performed within 2-3 minutes. In principle, a calibration stays valid as long as the MMF position/deformation and ambient conditions are unchanged. In practice, we found that the TM stays valid for more than an hour under our experimental conditions. After calibration, in order to perform imaging, the sample is translated back so
that the MMF illuminates a region with fluorescent beads and red blood cells (RBCs).

\section{Results}

\subsection{\label{sec:focus}Characterization of the focused spots}

\begin{figure}[b]
\centering
\includegraphics[width=\linewidth]{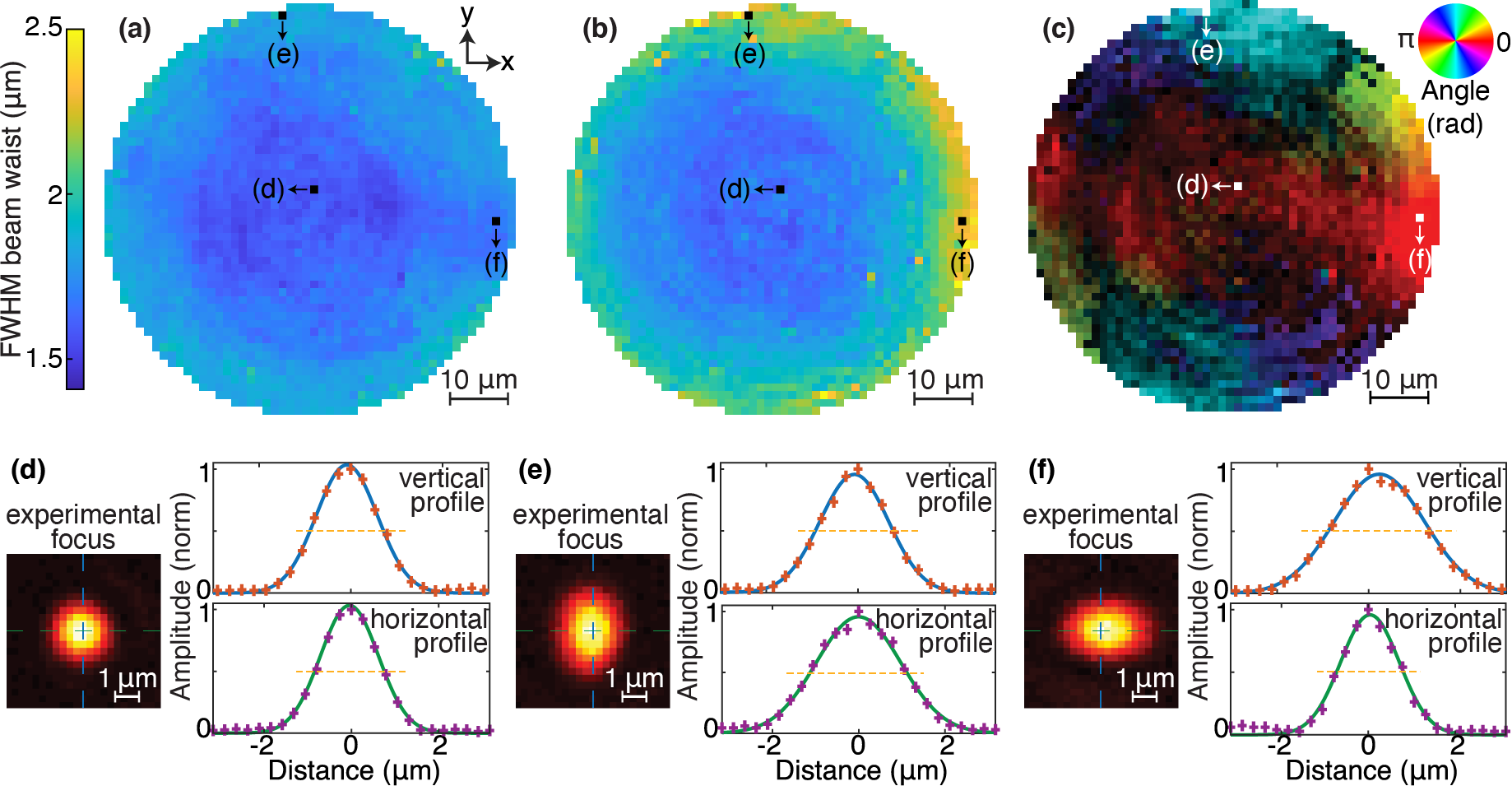}
\caption{(a-c)~Mapping of the FWHM and beam orientation of the focused spots as a function of their position. Values extracted from 2D Gaussian-like fit with (a)~the small axis $a_s$, (b)~the large axis $a_\ell$,  (c)~the orientation angle combined with a contrast scale bar proportional to  $r=(a_\ell-a_s)/(a_\ell+a_s)$. (d-f)~Experimental beam profile, experimental (dots) and fitted (lines) of the beam vertical horizontal profile at its maximum (dotted lines in the focus image).}
\label{fig:focus}
\end{figure}

We first provide a characterization of the focused spots obtained in the imaging plane (100~\textmu m away from the endoscope tip, where the sample is placed). The endoscope tip is immersed in a glass-bottom well, identical to the one used for the imaging experiments (Sec.~\ref{sec:imaging}), filled with distilled water but that does not contain fluorescent beads nor RBCs. An image of the spot
is acquired with the CMOS camera for every position in the scanned region (about 68~\textmu m diameter with a 1.3~\textmu m spacing). For each of the $\sim$2000 points, the experimental image of the focus is fitted to the following Gaussian-like profile:
\begin{equation}
F(x,y)=A \exp\left(-\frac{\left[\cos(\theta)x+\sin(\theta)y-x_0\right]^2}{2{\sigma_1}^2} -\frac{\left[-\sin(\theta)x+\cos(\theta)y-y_0\right]^2}{2{\sigma_2}^2}\right),
\label{eq:fit}
\end{equation}
where $x$ and $y$ are the two axis of the scan (see Fig.~\ref{fig:focus}(a)), $A$ is an amplitude term, $\theta$ is the orientation angle of the focus spot, $(x_0,y_0)$ are the coordinates of its fitted center, and $\sigma_{1,2}$ are the two (fitted) Gaussian RMS widths. The full width at half maximum (FWHM) of the focused spot in the two directions  is then  calculated: $a_{1,2}=2\sqrt{2\text{ln}(2)}\sigma_{1,2}$. For each spot, the smaller and larger widths are identified and labeled  $a_s$ and $a_\ell$, respectively. Fig.~\ref{fig:focus} presents the maps of $a_s$ (Fig.~\ref{fig:focus}(a)) and $a_\ell$ (Fig.~\ref{fig:focus}(b-c)), as well as the spot orientation ($\theta$, Fig.~\ref{fig:focus}(c)), for every point of the scanned area, as well as three typical examples of the beam profile in the center and in the area rim. In the orientation map, an intensity scale, proportional to the degree of anisotropy (estimated by $r=(a_\ell-a_s)/(a_\ell+a_s)$), is added to the colormap to darken the areas where the spot is more isotropic (i.e., when $r\to0$). 

From Fig.~\ref{fig:focus}(a-b), it appears that the beam profile is homogeneous and isotropic over a $\sim$50~\textmu m diameter area with $a_s=1.6\pm0.1$ and $a_L=1.7\pm0.1$~\textmu m. 
While the small radius remains globally constant over the whole scanned area, the large one increases sensitively at the area rim as it can be seen in Fig.~\ref{fig:focus}(b) as well as in  Fig.~\ref{fig:focus}(h,l), where $a_\ell/a_s$ varies up to 1.5. The orientation map (Fig.~\ref{fig:focus}(c) reveals that the spot is elongated along the radial direction as it approaches the cladding. 
This anisotropic elongation is caused by a decrease of the effective NA away from the optical axis. This occurs when the imaging plane is at some distance beyond the MMF end face (in our case 100~\textmu m). Indeed, when the focus is on the optical axis, the rays, emerging from the fiber tip, converge onto it in a symmetrical cone, but, when the focus moves away, this cone is cropped on one side by the edge of the fiber core. Hence, the effective NA is reduced in the radial plane and the spot becomes elongated.
Near the center of the field of view, the smallest spot size measured is 1.5~\textmu m. However, the NA of the incident beam 
(NA$\sim$0.25), lower than the nominal NA of the GRIN fiber (0.29), leads in first approximation to a diffraction-limited spot of FWHM $\lambda/(2\text{NA})\simeq1.1$~\textmu m. Even in the optical axis, the NA reduces with distance from the MMF tip. Since the NA decreases with the radial position in GRIN fibers, light rays emerging from the fiber at larger distances from the center will subtend a narrower range of angles. Considering a focus on the optical axis at some distance beyond the fiber end face, the effective NA would be limited by the more tilted rays emerging from the fiber that can reach the focus. For a plane at a distance of 100~\textmu m, a maximum NA of 0.25 and a fiber with perfectly parabolic index profile, the effective NA is estimated to be 0.19, leading to a spot FWHM of $\sim$1.4~\textmu m which is close to our experimental value. In the imaging experiments in the next section, the imaging diameter area is limited to $\sim$64~\textmu m to avoid the more anisotropic cases at the disc rim. 
\subsection{\label{sec:imaging}Imaging results}

\begin{figure}[t]
\centering
\includegraphics[width=\linewidth]{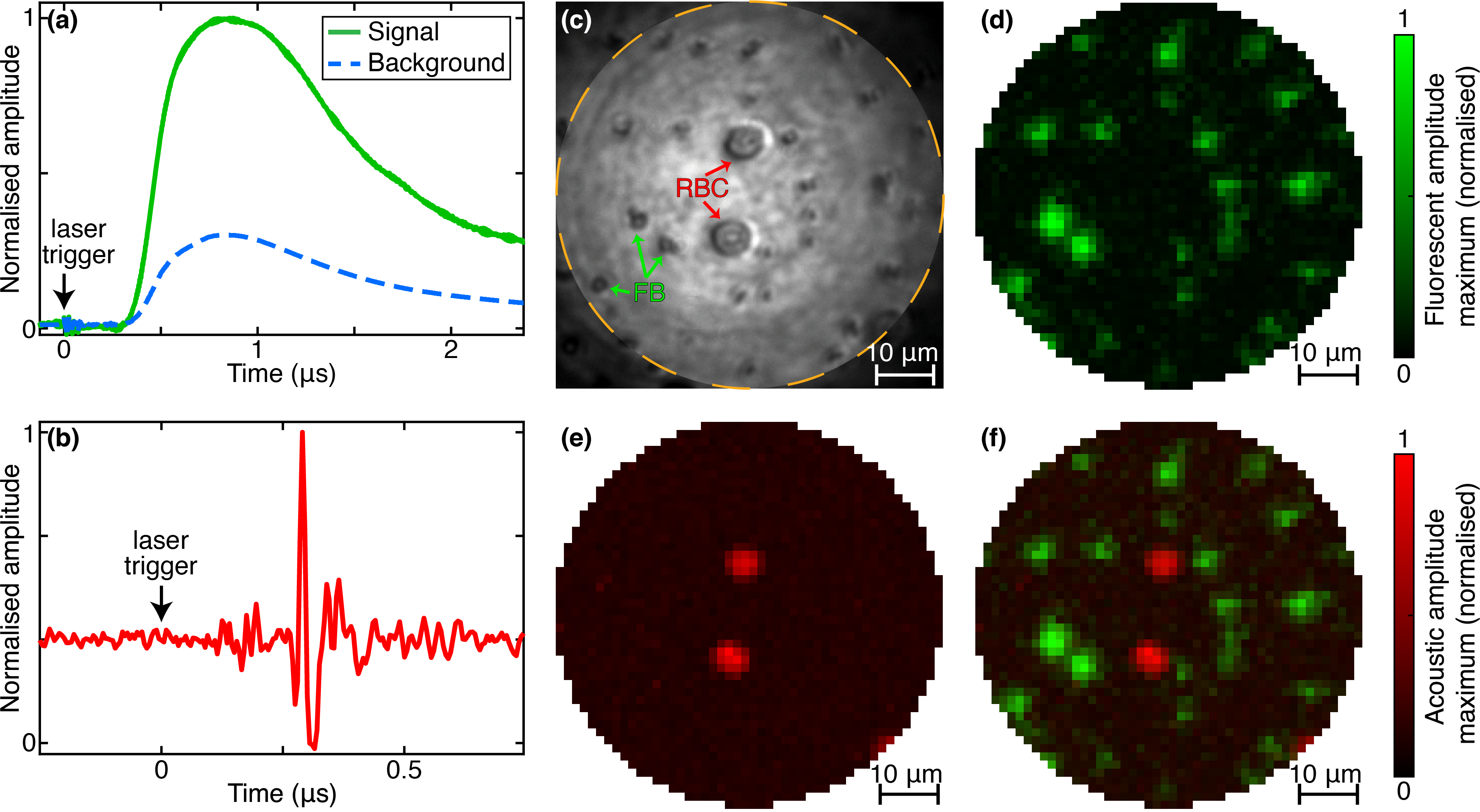}
\caption{(a)~Fluorescent signal acquired on a fluorescent bead and measured background in data. (b)~Acoustic signal acquired on a red blood cell. (a-b)~$t=0$ corresponds to the laser trigger detection. (c)~Bright-field image of the sample. RBC: Red Blood Cell, FB: Fluorescent Bead. Dotted line limits the scanned area. (d)~Fluorescent (e)~Photoacoustic and (f)~Composite image of the maximum value acquired with single-shot per pixel illumination.} 
\label{fig:results}
\end{figure}

Once the calibration is achieved, one can focus on a  RBC or a fluorescent bead to acquire a photoacoustic and a fluorescence signal, respectively, upon absorption of the 532~nm laser pulse.
Fig.~\ref{fig:results}(a) and (b) present a typical example of both these signals, generated after a single laser pulse. 
(The time $t=0$ indicates when the laser trigger is detected by the USB oscilloscope, but a delay is observed on both photoacoustic and fluorescent signals due to the amplifiers used in each channel. Moreover the fluorescence pulse is also lengthened by the amplifier at the output of the photomultiplier tube.)
Fig.~\ref{fig:results}(c) is a bright-field image of the raster-scanned area, acquired by the CMOS camera (see Fig.~~\ref{fig:setup}) with a incoherent light source. This proof-of-principle sample contains RBCs (two in the middle and a partially included one in the bottom right part) of $\sim$6~\textmu m in diameter and several 1~\textmu m-fluorescent beads (about twenty). 
The sample is scanned twice over $\sim$1800 points spaced by 1.3~\textmu m: a first time at low laser pulse energy ($E\simeq100$~nJ) for fluorescence imaging, and a second time at higher energy per pulse ($E\simeq2.10$~\textmu J) for photoacoustic imaging.
This double acquisition is required because of the difference between the excitation energy for fluorescence, which should be low enough to avoid optical saturation of the fluorophores and photobleaching, and the one required for single RBC to generate acoustic waves strong enough to be detected.  

We can observe that, thanks to the high power density in the focused spot and the high sensitivity to large angles afforded by the FOS detector, both fluorescence and photoacoustic signals display a high SNR.  
This confirms the possibility to perform fluorescence and photoacoustic endoscopy with a single-shot per pixel, as shown by the images of Fig.~\ref{fig:results}(d)-(f).

Figure~\ref{fig:results}(d) displays the image obtained by taking the maximum of the fluorescence signal acquired at each scan position after background subtraction. 
The background (shown in Fig.~\ref{fig:results}(a)) is estimated by averaging the signal in regions with no beads. Its origin is likely fluorescence from the fibers' polymer coating. Although we removed the coating of the MMF over its entire length, the single mode fiber (of the FOS) that is adjacent to the MMF was only stripped over 1-2 centimeters and the remaining polymer coating could be excited by light leaking from the MMF. Thus, this background signal was subtracted from each signal  before extracting the maximum value in Fig.~\ref{fig:results}(d). 
Figure~\ref{fig:results}(e) shows the photoacoustic image obtained by displaying the maximum absolute value of the signal measured by the FOS at each scan position. Single RBCs are clearly detected with a large SNR and a noticeable absence of background in this image. Obtaining a photoacoustic image with such an important SNR and in single-shot per pixel on single red blood cells demonstrates the potential of our new endoscope and is very promising for medical applications.
Finally, in Fig.~\ref{fig:results}(f) both fluorescent and photoacoustic signals are combined in an image showing the two modalities. The comparison with the brightfield image (Fig.~\ref{fig:results}(c)) exhibits a good agreement over the whole image.

From these results, the apparent size of the fluorescent beads and of the RBCs can be evaluated. As the former ones (1~\textmu m diameter) are smaller than the expected spatial resolution ($1.7\pm0.1$~\textmu m for the larger width, see Sec.~\ref{sec:calibration}), single bead images are fit with a 2D Gaussian function.  It leads to an estimation of the bead mean FWHM of $2.2$~\textmu m. This is in qualitative agreement with the expected value, although it is slightly larger. This difference could be attributed to a small defocus of the beads relative to the calibration plane, set for RBCs. The experimental FWHM of the latter can also be measured from this experiment; it yields a diameter varying from 5.8 to 6.1~\textmu m, in agreement with the expected size of RBCs.

\section{Conclusion}

A microendoscope with a footprint of only 250$\times$125~\textmu m$^2$  performing both photoacoustic and fluorescent imaging with a single-shot illumination per pixel  is demonstrated. It relies on a MMF to guide the excitation light and collect the fluorescent signal while an adjacent fiber-optic ultrasound sensor detects photoacoustic waves. 
Using the transmission matrix approach and a fast LC-SLM, the output field pattern can be controlled through wavefront shaping. We exploit this method to scan a focused spot in an imaging plane located 100~\textmu m beyond the MMF end face. 
This configuration necessarily induces inhomogeneities of the spot size and shape across the whole illuminated region. However, we managed to focus into a $1.7\pm0.3$~\textmu m FWHM spot throughout a circular area of $\sim$65~\textmu m diameter, allowing to perform raster-scan imaging. 
For each focus point, a single-shot laser excitation is enough to provide a fluorescence and a photoacoustic signal with a large SNR. We demonstrate imaging in both modalities by detecting multiple 1~\textmu m fluorescent beads and $\sim$6~\textmu m red blood cells, which could both be resolved individually.

To our knowledge, these results reports the first realization of a photoacoustic imaging with single-shot per pixel in minimally invasive endoscopy imaging. This is obtained by combining wavefront shaping techniques and a highly sensitive fiber-optic ultrasound sensor.
Single-shot acquisition 
brings on substantial improvement in terms of acquisition time. In the present setup, the latter was limited by the acquisition system (USB oscilloscope) which can sustain a maximum trigger rate of about 60~Hz, corresponding to the acquisition of one image in about 30 seconds. This system already exhibits the fastest acquisition rate attained in photoacoustic endoscopy, compared with current literature ($\sim$3~Hz in Ref.~\cite{stasio15}, and 22~Hz in Ref.~\cite{caravaca19}). A better acquisition card would allow performing similar imaging at 380~Hz (the LC-SLM refresh rate). At such speed, one image would be obtained in 5~seconds. A similar approach could also be achieved with a DMD which exhibits a 20~kHz switching rate. The acquisition would then only be limited by the laser repetition rate, which is 7~kHz in the present setup, leading to 4 images per second. Comparison of the images obtained with both modulator devices would also be of interest. 

This type of ultrathin endoscope, capable of fast photoacoustic and fluorescence imaging, is promising for \textit{in vivo} applications such as deep brain functional imaging, as it enables to monitor simultaneously and in real-time vascular dynamics (through the photoacoustic signal) and neuronal activity using fluorescence reporters.

\section{Acknowledgments}

This project has received funding from the European Research Council (ERC) under the European Union's Horizon 2020 research and innovation programme (ERC Consolidator Grant 681514 and ERC Advanced Grant 741149) and  Wellcome/EPSRC Centre for Surgical and Interventional Sciences (Ref. NS/A000050/1). A.M. Caravaca-Aguirre was supported by the Marie Sk\l{}odowska-Curie Individual Fellowship (Project DARWIN, Grant Agreement No. 750420).


\bibliography{main}

\end{document}